\documentstyle[prl,preprint,epsfig,aps]{revtex}
\draft
\tighten

\begin{document}

\title{Fusion of light proton-rich exotic nuclei at near-barrier energies}
\author{P. Banerjee, K. Krishan, S. Bhattacharya and C. Bhattacharya}
\address{Physics Group, Variable Energy Cyclotron Centre,
\\1/AF Bidhan Nagar, Kolkata - 700 064, INDIA}
\date{\today}
\maketitle

\begin{abstract}
We study theoretically fusion of the light proton-rich exotic nuclei
$^{17}$F and $^8$B at near-barrier energies in order to investigate the
possible role of breakup processes on their fusion cross sections. To this 
end, coupled channel calculations are performed considering the couplings
to the breakup channels of these projectiles. In case of $^{17}$F, the coupling
arising out of the inelastic excitation from the ground state to the bound
excited state and its couplings to the continuum have also been taken into
consideration. It is found that the inelastic excitation/breakup of $^{17}$F 
affect the fusion cross sections very nominally even for a heavy target like
Pb. On the other hand, calculations for fusion of the one-proton halo nucleus 
$^8$B on a Pb target show a significant suppression
of the complete fusion cross section above the Coulomb barrier. 
This is due to the larger breakup probability of $^8$B as compared to that of
$^{17}$F. However, even for $^8$B, there is little change in 
the complete fusion cross sections as compared to the no-coupling case at 
sub-barrier energies.
\end{abstract}


\section{Introduction}
The increasing availability of radioactive ion beams has made possible the 
study of the interactions and structure of exotic nuclei far from the line
of stability. Nuclei located near the neutron or proton drip 
lines, for which the valence particles are very loosely bound, give rise to 
interesting new phenomena, e.g., formation of halo structures \cite{han95}. 
The low binding energies of the valence nucleons in these nuclides result 
in large root mean 
square (rms) radii and thus, in increased probabilities for specific reaction 
channels such as breakup, nucleon tranfer and fusion. Breakup reactions, in
which the valence nucleons are removed from these exotic nuclei, have very
large cross sections of a few barns. This is particularly true for the 
neutron-halo nuclei with well-developed halo structure, e.g., $^{11}$Be, 
$^{11}$Li. However, the breakup cross sections are quite significant also for 
a light one-proton halo nucleus like $^8$B \cite{moto}.

Since a large flux goes into the breakup channel, many questions
concerning the effects of breakup processes on sub-barrier fusion of 
drip-line nuclei have been raised recently, both from the experimental 
\cite{taka97,rehm,sign,kola,das99,trot} and theoretical 
\cite{huss,taki,dasso94,hagi00} points of view. The presence of halo 
structure in these exotic nuclei means a rms matter radius larger than the 
usual value deduced from the 
systematics. Consequently, the sub-barrier fusion cross section should
be enhanced since the Coulomb barrier is lowered. On the other 
hand, one might think that increased breakup probabilities
for these nuclei would remove a significant part 
of flux and thus cross sections for complete fusion would be hindered. 

Very recently fusion at near-barrier energies has been 
investigated theoretically for the neutron halo nuclei $^{11}$Be and $^6$He
on Pb. Coupled channel calculations have been performed by 
discretizing in energy the particle continuum states \cite{hagi00}. The 
calculations show that the couplings to the breakup channels have appreciable
effects. At energies above the Coulomb barrier, it is found that the cross 
sections for complete fusion are hindered compared to the no-coupling case. On 
the other hand, at sub-barrier energies the complete fusion cross sections are 
enhanced significantly. Recent measurements on fusion cross sections for the
loosely bound nucleus $^9$Be \cite{das99} and the two-neutron halo nucleus 
$^6$He \cite{trot} on heavy targets at near-barrier energies are in general 
agreement with the theoretical findings in Ref.~\cite{hagi00}. 

However, for the proton-rich systems, the effect of coupling to the breakup
channels on the fusion process may not be as significant as the neutron-rich
nuclei. This is because unlike the neutron-rich systems
the valence proton in the loosely bound proton-rich exotic nuclei has
to tunnel through the barrier resulting from the Coulomb repulsion due
to the charged core, which hinders the formation of halo. In fusion
reaction of such a nucleus with a target, this might not lead to a 
significant lowering of the Coulomb barrier. 

Fusion reaction induced by a proton-rich nucleus has received attention 
only recently. First measurement of fusion cross sections has been done
for $^{17}$F \cite{rehm}. Therefore, we present 
results of calculations on fusion cross sections
of the proton drip-line exotic nucleus $^{17}$F
and compare with the data. But it would also be quite
interesting to study fusion induced by a proton-rich nucleus with large
breakup probabilities, e.g., the one-proton halo nucleus $^8$B. 
In this paper, we also report first calculations on fusion reactions 
induced by the one-proton halo nucleus $^8$B. We
follow the same theoretical formalism as in Ref.~\cite{hagi00}. Within
this coupled channel method of calculation, we neglect continuum-continuum 
couplings. We assume the target to be inert in each case, i.e., possible 
target excitations are neglected. We choose a heavy target because
it would be ideal to study the effects of inclusion of coupling to the
breakup channels, as breakup effects increase with target mass.

In section 2, we present the theoretical formalism. Results and discussions
are given in section 3. Finally, we summarize and conclude in section 4.

\section{Formalism}
The coupled channel equations for the projectile($P$)-target($T$) system 
(with reduced mass $\mu$) in 
the isocentrifugal approximation \cite{hagi95} are given by \cite{hagi99}
\begin{eqnarray}
\big[-{\hbar ^2\over 2\mu}{d^2\over dr^2}+{l(l+1)\hbar ^2\over 2\mu r^2}
+V^{(0)}_N(r)+{Z_PZ_Te^2\over r}+\epsilon _n - E\big]\psi _n(r)
=-\sum_mV_{nm}(r)\psi _m(r).
\end{eqnarray}
In writing the above equations, the angular momentum of
the relative motion in each channel has been replaced by the total angular
momentum $l$ \cite{hagi95}. In eqs.(1), $V^{(0)}_N$ is the nuclear potential
in the entrance channel and $\epsilon _n$ is the excitation energy of
the $n-$th channel. Here we assume that $n=0$ labels the ground state of the
projectile nucleus and all other $n$'s refer to bound excited states or
particle continuum states of the projectile. $V_{nm}$ are the coupling form
factors computed on the microscopic basis. They are obtained by folding the
external nuclear and Coulomb fields with the proper single-particle 
transition densities, in which the last weakly bound nucleon is promoted
from the ground state to excited states below the breakup threshold or in
the continuum. More explicitly, the coupling form factor for the promotion
of the single valence particle from the bound state $(n_1l_1j_1m_1)$ to
the continuum state $(l_2j_2m_2)$ with continuum energy $E_c$ is given by
\begin{eqnarray}
V_{nm}(r)&=&\sum_{m_c}\langle j_1m_1j_cm_c\mid J_iM_i\rangle\langle j_2m_2
j_cm_c\mid J_fM_f\rangle\nonumber\\&\times& f_{n_1l_1j_1m_1\rightarrow
l_2j_2m_2}(r),
\end{eqnarray}
where $J_{i(f)}, M_{i(f)}$ are the total spin of the  projectile and its
projection quantum number in the initial (final) state. $j_c$ is the
spin of the core and $m_c$ is its projection. We have assumed a {\it core +
single particle} structure for the projectile.
The single-particle form factor is given by
\begin{eqnarray}
&&f_{n_1l_1j_1m_1 \rightarrow l_2j_2m_2}(r)\nonumber\\&=&\sqrt{\pi}
\sum_{\lambda}(-1)^{m_2
+{1\over 2}}\delta(l_1+l_2+\lambda,even)\langle j_1{1\over 2}j_2-{1\over 2}
\mid \lambda 0\rangle\nonumber\\&\times&{\sqrt{2j_1+1}\sqrt{2j_2+1}\over 
\sqrt{2\lambda+1}}\langle j_1-m_1j_2m_2\mid \lambda (m_2-m_1)\rangle\sqrt{
2\lambda+1\over 4\pi}\delta_{m_1,m_2}\nonumber\\&\times&\big[\int ^{\infty}
_0r'^2dr'\int ^{+1}_{-1}du R^{\star}_{E_cl_2j_2}(r')R_{n_1l_1j_1}(r')
V_T(\sqrt{r^2+r'^2-2rr'u})P_{\lambda}(u)\big],
\end{eqnarray}
where $P_{\lambda}$ is the Legendre polynomial. The functions 
$R_{E_cl_2j_2}(r)$ and $R_{n_1l_1j_1}(r)$ are the radial parts of the 
single-particle wave functions for the continuum and bound states, 
respectively. In case of a 
transition from the ground state to a bound excited state, $R_{E_cl_2j_2}(r)$
would be replaced by $R_{n_2l_2j_2}(r)$ in eq.(3). The potential $V_T$,
involving both a nuclear and a Coulomb component, is the mean field experienced
by the single particle due to the presence of the target, responsible for
the transition. 

The coupled channel eqs.(1) are solved by imposing the incoming wave
boundary condition (IWBC) \cite{land}, where there are only incoming waves 
at $r_{\mbox {\rm min}}$ which is taken to be the minimum position of the 
Coulomb pocket inside the barrier. This type of boundary condition is valid
for heavy-ion reactions, for which there is a strong absorption inside the
Coulomb barrier. The boundary conditions are expressed as
\begin{eqnarray}
\psi _n(r)&\rightarrow&T_n\exp(-i\int ^r_{r_{\mbox{\rm min}}}k_n(r')dr'),~~~~
r\leq r_{\mbox{\rm min}},\\&\rightarrow&H_l^{(-)}(k_nr)\delta _{n,0}+R_n
H_l^{(+)}(k_nr),~~r>r_{\mbox{\rm max}},
\end{eqnarray}
where
\begin{eqnarray}
k_n(r)=\sqrt{{2\mu\over \hbar ^2}(E-\epsilon _n-{l(l+1)\hbar ^2\over 2\mu
r^2}-V^{(0)}_N(r)-{Z_PZ_Te^2\over r})}
\end{eqnarray}
is the local wave number for the $n-$th channel and $k_n$, in Eq.(5), is
equal to $\sqrt{2\mu(E-
\epsilon _n)/\hbar ^2}$. $H_l^{(-)}$ and $H_l^{(+)}$ are the 
incoming and
outgoing Coulomb functions, respectively. $r_{\mbox{\rm max}}$ is taken to
be a distance beyond which both the nuclear potential and the Coulomb 
coupling are sufficiently small.

The fusion probability is defined as the ratio between the flux inside the
Coulomb barrier and the incident flux. For the boundary conditions given by 
eqs.(4) and (5), it becomes
\begin{eqnarray}
P_n={k_n(r_{\mbox{\rm min}})\over k_0}\mid T_n\mid^2
\end{eqnarray}
for the $n-$th channel. Complete fusion is a process where all the nucleons 
of the projectile are captured by the target nucleus. We thus define cross
section of complete fusion using the flux for the non-continuum channel
(i.e. $n=0$) as \cite{hagi98}
\begin{eqnarray}
\sigma _{\mbox{\rm CF}}={\pi\over k^2_0}\sum _l(2l+1)P_0=
{\pi\over k^2_0}\sum _l(2l+1){k_0(r_{\mbox{\rm min}})\over k_0}\mid T_0
\mid^2.
\end{eqnarray}
The flux for the particle continuum channel ($n\neq 0$) is associated with 
incomplete fusion, whose cross section is given as
\begin{eqnarray}
\sigma _{\mbox{\rm ICF}}={\pi\over k^2_0}\sum _l(2l+1)\sum_{n\neq 0}P_n=
{\pi\over k^2_0}\sum _l(2l+1)\sum_{n\neq 0}{k_n(r_{\mbox{\rm min}})\over k_0}
\mid T_n\mid^2.
\end{eqnarray}
Eqs.(8) and (9) are correct when there is only one bound state below the
breakup threshold. If there are some bound excited states below the threshold
other than the ground state, the summations in these equations are to be 
carried out accordingly.

\section{Results and discussions}
In recent measurements with the proton drip-line exotic nucleus $^{17}$F in 
fusion-fission reaction on Pb at energies around the Coulomb barrier no 
enhancement of the fusion cross section is observed \cite{rehm}. In Fig.~1,
we show our calculations on the fusion cross sections for $^{17}$F + Pb at
near-barrier energies alongwith the data \cite{rehm}. In our calculation, we
have considered the effect of the
transition from the ground state of $^{17}$F (600 keV below the breakup 
threshold) into the continuum. The bound excited state
situated 495 keV above the ground state is also included in the calculation.
This means that we have considered ground state (0$d_{5\over 2}$) to first 
excited state (1$s_{1\over 2}$) coupling (through quadrupole transition) and
also couplings from the first excited state to the continuum. 
The continuum up to 2 MeV has been considered and it has been found
that the results converge. The continuum has been discretized into 10 bins with
a bin size of 200 keV. The continuum states have been taken to be situated at
the middle of each bin. We consider transitions of multipolarity 1 and 2 
for transitions into the continuum, with $s$-, $p$-, $d$- and $f$-waves in
the continuum. The nuclear part of the valence 
proton-target interaction, causing the inelastic transition/breakup of 
$^{17}$F, has been taken to be the same as 
the neutron-target interaction in Ref.~\cite{hagi00}. The nuclear part of the
ion-ion interaction potential has been taken to be equal to that of the 
neighbouring nucleus $^{16}$O on Pb \cite{vid}. As far as the structure of 
$^{17}$F is concerned, we assume a single particle potential model in which the
valence proton moves in a Coulomb and Woods-Saxon potential relative to $^{16}$O
core. The depth of the Woods-saxon potential has been adjusted to reproduce 
the known one-proton separation energies for the bound states. 
The radius and diffuseness parameters of the Woods-Saxon potential have been 
taken to be 1.25 {\it fm} and 0.5 {\it fm} respectively. This gives a rms proton
radius of 3.7 {\it fm}.

There is good overall agreement of our calculations with the data. But
compared to the no-coupling case, there is no discernible change 
of the fusion cross sections when the channel coupling effects
are considered (see Fig.~1). 

However, observations are quite different when calculations on the fusion
cross sections of $^8$B on the same target (Pb) have been done. We use the 
single-particle potential model used by Esbensen and Bertsch to describe the
structure of $^8$B \cite{esb,pb}. This yields a rms proton radius of 4.24 
{\it fm}. We consider transitions into the continuum of 
multipolarities 1 and 2, and include $J$ = 1, 2 and 3 final channels in the
couplings. $s$-, $p$-, $d$- and $f$-waves in the continuum are considered for 
the appropriate transitions. The inclusion of the $J$ = 4 final channel in the 
calculation has almost got no effect. The $J$ = 0 final channel has been left 
out of calculation as the 
$0^+$ final state appears to be very weak in the low-lying excitation spectrum
of $^8$B \cite{esb}. The continuum discretization scheme is the same as that 
for $^{17}$F. The 1$^+$ and 3$^+$ resonances at 0.637 MeV and 2.183 MeV have
also been included in the calculations. The nuclear potential for the
$^8$B + Pb system has been taken to be equal to that between the neighbouring
nucleus $^7$Be and Pb as in Ref.~\cite{das99}.  As for $^{17}$F, the nuclear 
part of the valence proton-target interaction, causing the inelastic 
transition/breakup of $^8$B, has been taken to be the same as 
the neutron-target interaction in Ref.~\cite{hagi00}. 

The results are displayed in Fig.~2. At sub-barrier energies, 
there is some 20\% increase of the total (complete + incomplete) fusion
cross sections (dashed line) compared to the zero coupling case. But results 
of channel coupling are nominal so far as complete fusion cross sections (thick
solid line) in this region
are concerned. However, at energies above the Coulomb barrier, there
is significant suppression of the complete fusion cross sections. The 
complete fusion cross section is $\sim$69\% of the total fusion cross section,
for example, at 60 MeV centre-of-mass (c.m.) energy. This
feature of fusion cross sections in $^8$B induced reaction on a heavy target
at above-barrier energies is the same as that for the one-neutron halo
nucleus $^{11}$Be. But compared to the appreciable enhancement
of the complete fusion cross sections in case of $^{11}$Be \cite{hagi00,pb1}, 
the changes observed are small in the sub-barrier region.

If only dipole transitions are considered, the breakup effects will be 
smaller compared to the full (dipole + quadrupole) calculations and we
expect that the suppression of the complete fusion cross sections will be
less. For $^{17}$F + Pb reaction, practically no difference is observed
as to this difference in calculation. However, for the $^8$B induced reaction
on Pb, this effect is visible at above-barrier energies (see Fig.~3). But 
the increase in the complete fusion cross section is not much, e.g., $\sim$4\% 
at 60 MeV. This indicates that the 
quadrupole breakup contributions are small in case of $^8$B + Pb reaction. 

The breakup probabilities of an
exotic nucleus depend significantly on the separation
energy of the valence nucleon and its orbital angular momentum 
configuration \cite{raja}. Any non-zero angular momentum with respect to
the core will lead to a centrifugal barrier, which will restrict the
extent of the wave function in the coordinate space. This, in turn, 
decreases the cross sections of breakup processes in which the valence 
particle is removed from the nucleus \cite{raja}. Increase of 
separation energy of the valence nucleon further decreases the breakup 
probabilities \cite{raja}.

The dominant ground state
configuration of $^8$B is $^7$Be$({3\over 2}^-)\otimes \pi$(0$p_{3\over 2}$),
whereas for $^{17}$F it is $^{16}$O$(0^+)\otimes \pi$(0$d_{5\over 2}$). The 
valence proton in $^{17}$F feels more than two times the Coulomb barrier as the 
last 
proton in $^8$B. The $\ell$=2 centrifugal barrier experienced by the valence
proton in $^{17}$F is also almost 3.5 times larger compared to the $\ell$=1 
centrifugal 
barrier for the valence proton in $^8$B. The one-proton separation energy 
(0.6 MeV) in $^{17}$F is almost 4.5 times that (0.137 MeV) in $^8$B. Thus, 
breakup through one-proton removal is much more favoured in case of $^8$B 
compared to $^{17}$F. Dynamical calculations in Ref.~\cite{rehm} also
show that the chances of inelastic excitation/breakup of $^{17}$F are small.
On the other hand, the breakup probabilities are quite significant for $^8$B
on a heavy target \cite{moto}. In fact, within a semiclassical theory 
\cite{bert}, we compute the breakup cross section 
of $^8$B on Pb at 60 MeV in the laboratory system 
arising out of electric dipole transition 
to be as large as 5.1 b. This is close to 4.6 b, the $E1$ dissociation 
cross section for the one-neutron halo nucleus $^{11}$Be on the same target at 
the same incident energy. On the other hand, for $^{17}$F on Pb target at
near-barrier energy, the cross section for the same process comes out to 
be 46.46 mb only. Therefore, it is expected that compared to $^{17}$F, 
the breakup process would have larger effects on the fusion cross sections 
in $^8$B induced reactions.

\section{Summary and conclusions}
In summary, we have performed coupled channel calculations of fusion cross
sections in reactions induced by the proton-rich exotic nuclei $^{17}$F
and $^8$B on Pb target at near-barrier energies. Couplings to the inelastic
and/or breakup channels have been taken into consideration. For $^{17}$F
induced reaction, the channel coupling effects are minimal. However, for
$^8$B induced reaction, there is appreciable suppression of the complete
fusion cross sections at energies above the Coulomb barrier. The difference
in the features of the fusion cross sections in reactions involving these
two nuclei is attributed to the difference in their ground state 
configurations. $^8$B is a one-proton halo nucleus with larger breakup 
probabilities, most of its breakup cross sections originating due to dipole 
transitions.
On the other hand, $^{17}$F, although proton-rich, is close
to an ordinary nucleus so far as its size and breakup cross sections are 
concerned. We suggest measurement of the different components
of the fusion cross sections at near-barrier energies in reaction involving 
$^8$B and a heavy target.

One of the authors (P.B.) thanks J. N. De for his encouragement during this
work. Fruitful discussions with Andrea Vitturi are gratefully acknowledged.

\newpage
\begin{figure}
\begin{center}
\mbox{\epsfig{file=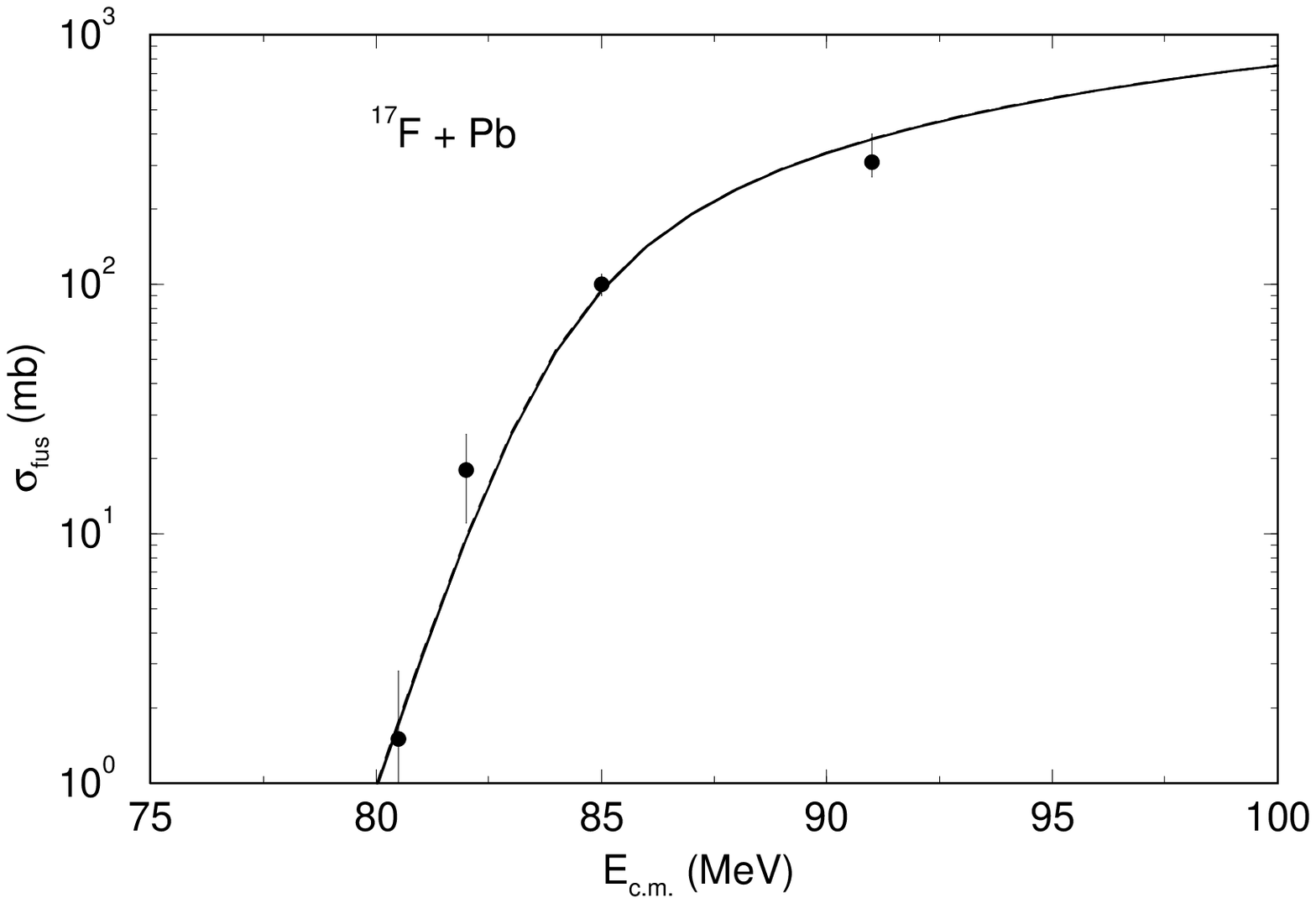,height=11cm}}
\end{center}
\caption{Fusion cross sections in the $^{17}$F + Pb reaction at near-barrier
energies. The thin solid line gives the cross sections with zero
couplings. The thick solid line and the dashed line give the complete and the
total (complete + incomplete) fusion cross sections respectively. The data
have been taken from Ref.~\protect\cite{rehm}.}
\label{fig:figa}
\end{figure}

\begin{figure}
\begin{center}
\mbox{\epsfig{file=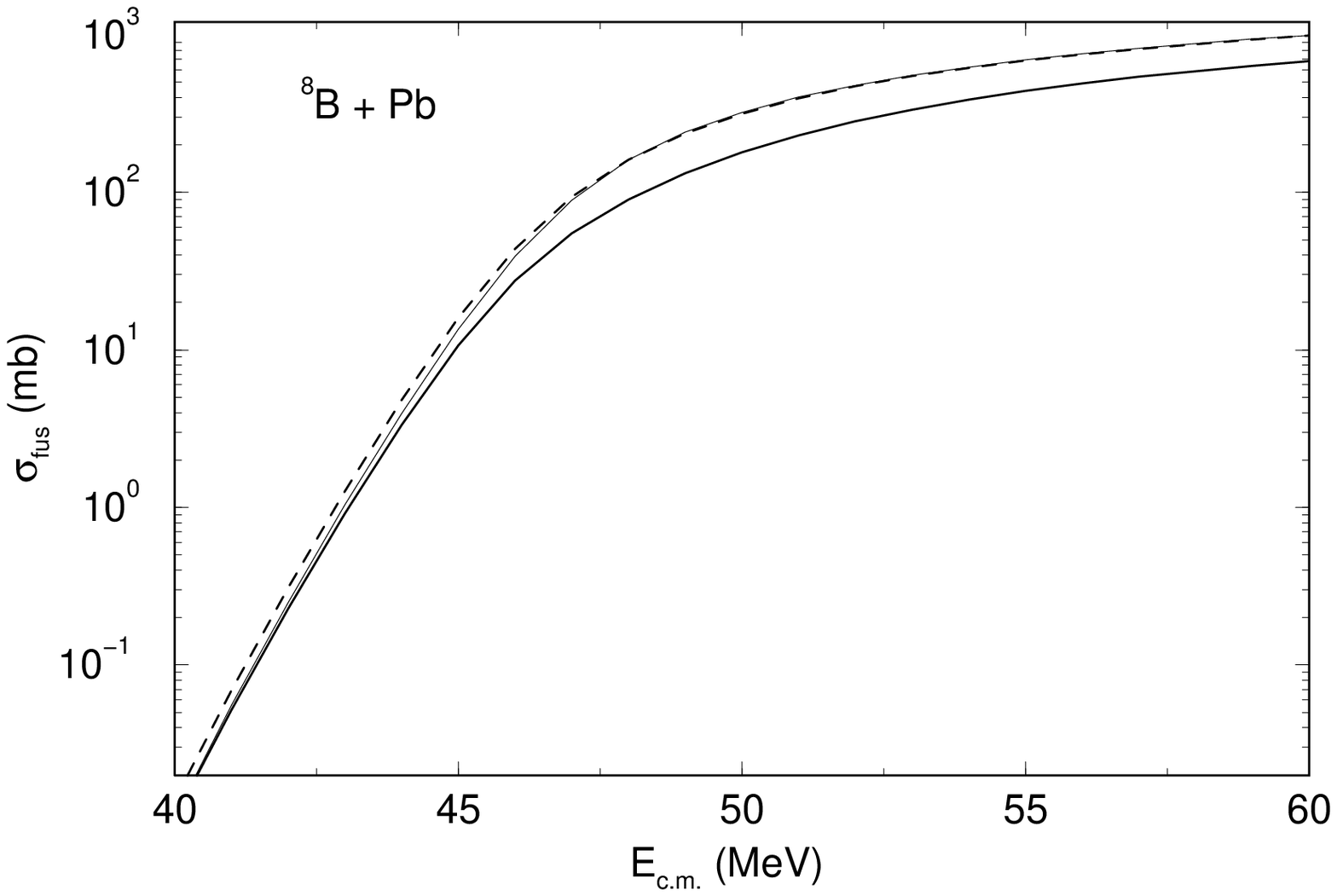,height=11cm}}
\end{center}
\caption{Fusion cross sections in the $^8$B + Pb reaction at near-barrier
energies. The thin solid line gives the cross sections with zero
couplings. The thick solid line and the dashed line give the complete and the
total (complete + incomplete) fusion cross sections respectively.}
\label{fig:figb}
\end{figure}

\begin{figure}
\begin{center}
\mbox{\epsfig{file=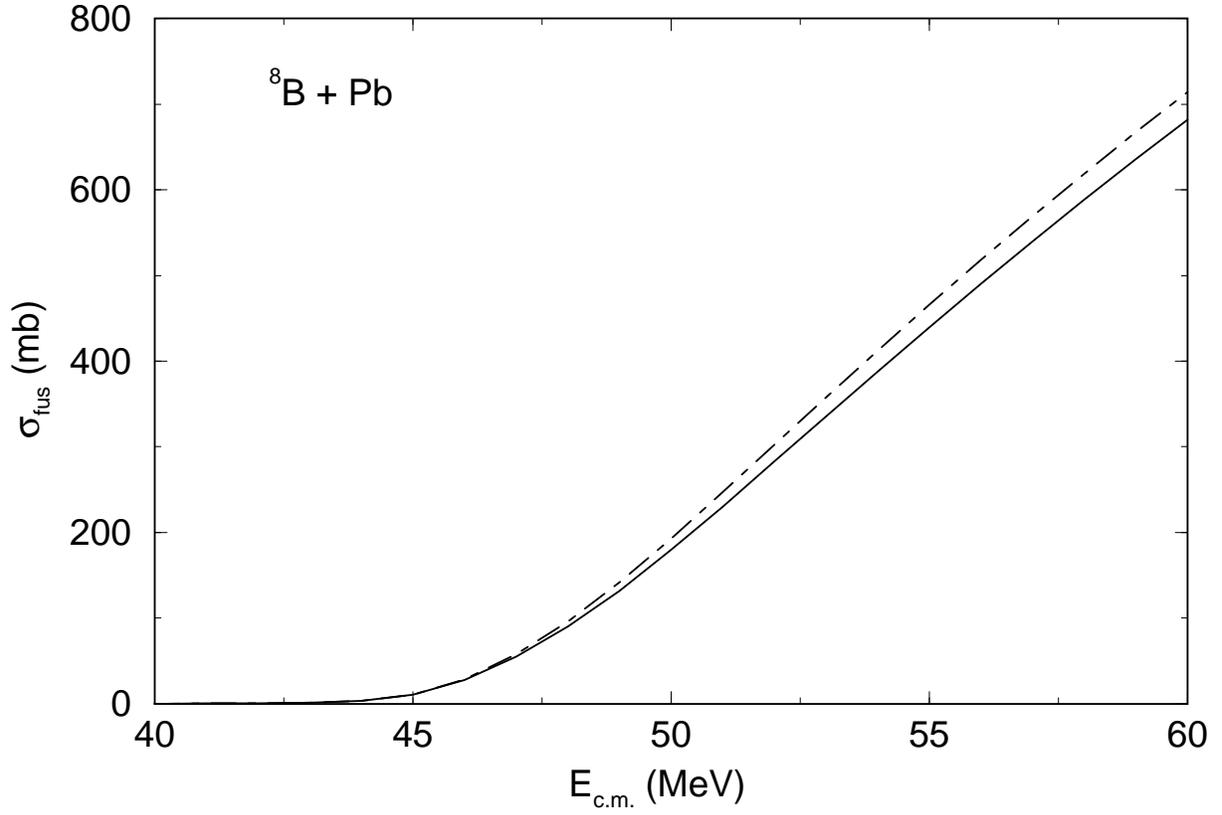,height=11cm}}
\end{center}
\caption{Complete fusion cross sections in the $^8$B + Pb reaction at 
near-barrier
energies. The dot-dashed line gives the cross sections for
dipole transitions only. The solid line gives the same for
dipole and quadrupole transitions.}
\label{fig:figc}
\end{figure}
\end{document}